\documentclass{nature-figs}

\usepackage{amsmath}
\usepackage{graphicx}
\usepackage[usenames]{color}
\usepackage{verbatim}

\def\Lmax{L_\text{max}}

\title{Orientational Order of Motile Defects in Active Nematics}

\author{Stephen J. DeCamp$^{1\ast}$, Gabriel S. Redner$^{1\ast}$,
  Aparna Baskaran$^1$, Michael F. Hagan$^{1}$ \& Zvonimir Dogic$^{1}$}

\begin{document}

% Single-space for easier reading.  Remove before submitting.
\spacing{1}

\maketitle

\begin{affiliations}
\item[$^1$] Department of Physics, Brandeis University, Waltham MA 02454, USA
\item[$^{\ast}$] These authors contributed equally to the work
\end{affiliations}

% Code for adding a footer containing the source information.  Remove
% this when generating for submission
%\thispagestyle{fancy}
%\pagestyle{fancy}

\begin{abstract}
The study of liquid crystals at equilibrium has led to fundamental
insights into the nature of ordered materials, as well as to practical
applications such as display technologies. Active nematics are a
fundamentally different class of liquid crystals, driven away from
equilibrium by the autonomous motion of their constituent rod-like
particles\cite{Narayan2007, Duclos2014, Zhou2014, Brugues2014}. This
internally generated activity powers the continuous creation and
annihilation of topological defects, which leads to complex streaming
flows whose chaotic dynamics appear to destroy long-range
order\cite{AditiSimha2002, Sanchez2012, Giomi2013a, Thampi2013c,
  Thampi2014a, Gao2014, Ngo2014}. Here, we study these dynamics in
experimental and computational realizations of active nematics. By
tracking thousands of defects over centimetre-scale distances in
microtubule-based active nematics, we identify a non-equilibrium phase
characterized by system-spanning orientational order of defects. This
emergent order persists over hours despite defect lifetimes of only
seconds. Similar dynamical structures are observed in coarse-grained
simulations, suggesting that defect-ordered phases are a generic
feature of active nematics.
\end{abstract}

Topological defects play important roles in diverse phenomena ranging
from high-energy physics and cosmology to traditional condensed matter
systems\cite{Chuang1991}. For instance, the spontaneous unbinding of
dislocation pairs mediates the melting of 2D
crystals\cite{Nelson1979}. Despite their usual role as centers of
disorder, defects can also organize into higher-order equilibrium
structures with emergent properties, such as liquid crystalline
twist-grain-boundary phases and flux-line lattices in
superconductors\cite{Renn1988, Brandt1995}. Far less is understood
about the role of defects in active matter systems, which are driven
away from equilibrium by the motion of their constituent particles
\cite{Toner1995, Toner2005, Vicsek2012, Marchetti2013, Palacci2013a,
  Redner2013, Weber2014, Wensink2012a, Tjhung2012a}. Previous work on
active nematics has demonstrated an instability at large
wavelengths\cite{AditiSimha2002} which leads to spontaneous defect
nucleation and unbinding\cite{Sanchez2012, Giomi2013a, Thampi2013c,
  Thampi2014a, Gao2014}. In contrast to the well-studied passive
defects found in equilibrium matter, defects in active nematics are
motile\cite{Keber2014}, and are continuously generated and
annihilated, producing a dynamical defect-riddled phase that is
inherently non-equilibrium. The observed dynamics are complex and
chaotic, and appear to destroy the long-range ordering of the
underlying nematic. Here, by tracking thousands of defects over long
times, we demonstrate that defects self-organize into a higher-order
phase with broken rotational symmetry. The orientational ordering of
defects spans macroscopic samples and persists for the sample lifetime
of many hours, despite the lifetimes of the constituent defects being
orders of magnitude shorter.

Our experimental system is comprised of micron-long stabilized
microtubules (MTs), streptavidin clusters of biotin-labeled kinesin
motors\cite{Nedelec1997} and the non-adsorbing polymer polyethylene
glycol (PEG) (Fig. 1a). In a bulk suspension, PEG induces formation of
MT bundles by the depletion
mechanism\cite{Needleman2004,Hilitski2015}. The same interaction also
depletes MTs onto a surfactant-stabilized oil-water
interface. Centrifugation makes it possible to spin down all the MT
bundles onto the interface, leading to the formation of a dense
quasi-2D MT film which exhibits local orientational order. Each
kinesin cluster binds to multiple MTs. As each motor within the
cluster hydrolyzes adenosine triphosphate (ATP), it moves towards the
plus end of a MT and induces inter-filament
sliding\cite{Hentrich2010}. This generates extensile mechanical
stresses that drive the nematic film away from equilibrium
(Fig. 1b). A biochemical regeneration system maintains a constant ATP
concentration and powers the system for over 24 hours (see
Supplementary Methods). We image these active nematics with both
fluorescence microscopy and LC-PolScope\cite{Shribak2003}. LC-PolScope
measures the orientation of the nematic director $\theta(\mathbf{r})$
with pixel resolution.  It also measures the magnitude of
birefringence which reveals the effective thickness of the nematic
film, or retardance $\Delta(\mathbf{r})$ (see Supplementary Figure 4
for an extended discussion).  Using a 4x objective, we observe the
dynamics of the MT film over the entire field of view, corresponding
to an area of 2.3$\times$1.7 mm.

In parallel, we have developed a tractable coarse-grained
computational model. Our approach employs Brownian dynamics
simulations of rigid spherocylinders which, in equilibrium, form a
nematic phase\cite{Bates2000}. Long-ranged hydrodynamic interactions
are omitted, producing an essentially dry system. The length of each
constituent rod increases at a constant rate, producing an extensile
stress similar to the motor-driven extension of MT bundles
(Fig. 1d). Upon reaching a pre-set maximum length, a spherocylinder is
split in half and two other rods are simultaneously merged, thus
keeping the total particle number fixed (see Supplementary Methods).
Though inspired by the dynamics of MT bundles, this approach is not
meant to quantitatively reproduce specific features of the
experimental system, but simply to capture its microscopic symmetries
(nematic interparticle alignment, and extensile nematic activity).

In equilibrium, nematic defects anneal to minimize free energy,
eventually producing a uniformly aligned state. It is not possible to
prepare an analogous state in extensile active nematics, since uniform
alignment is inherently unstable to bend
deformations\cite{AditiSimha2002}. These distortions grow in amplitude
and produce a fracture line, terminated at one end by a defect of
charge $+1/2$, and by a $-1/2$ defect at the other (Fig. 1c). The
asymmetry of $+1/2$ defects causes motor-generated stresses to produce
a net propulsive force, leading to extension of the fracture
line. Above a critical length, the fracture line self-heals, leaving
behind a pair of isolated, oppositely charged
defects\cite{Thampi2014a}. For experimental ATP concentration, $+1/2$
defects move at speeds of $\sim$8 $\mu$m/sec. Defects are transient
objects; on average, a $+1/2$ defect exists for 40 sec before
colliding with a $-1/2$ defect and annihilating, leaving behind a
uniformly aligned nematic region\cite{Shi2013} (Fig. 1c). The system
reaches a steady state in which the rate of defect generation is
balanced by the rate of annihilation (Supplementary videos 1-2). Very
similar patterns of defect generation and annihilation are observed in
the simulations, despite the fact that our computational model does
not include hydrodynamic interactions (Fig. 1e, 1f, Supplementary
videos 4-5).

We have developed algorithms that identify defect positions and
orientations from either LC-PolScope images or simulation
configurations, and track their temporal dynamics over the entire
lifetime of either a microtubule sample or simulation (see
Supplemental Information). Defect positions and orientations are
determined by measuring the winding of the local director. We define
the orientation of each comet-like $+1/2$ defect by drawing an arrow
from the comet's head to its tail (see Supplemental
Information). These algorithms allow us to analyze statistically large
defect populations, providing invaluable insight into their
self-organization at macroscopic scales.

It has been commonly assumed that the dynamics of motile defects leads
to a disordered chaotic state. However, quantitative analyses reveal
that this is not the case. We find the that orientational distribution
function of $+1/2$ defects is not flat, but exhibits two well-defined
peaks, implying the existence of a higher-order dynamical phase
(Fig. 2e). Although $+1/2$ defects are polar objects, we find that
they form a nematic phase in which they are equally likely to point in
either direction along the preferred axis (Fig. 2b). Furthermore, the
orientational distribution function of $-1/2$ defects exhibits
six-fold symmetry, although the strength of this order is
significantly less than the nematic order of $+1/2$ defects.
Additionally, radial distribution functions of both $\pm1/2$ defects
reveal no long-range positional order (Supplementary Fig. 9).

Next, we investigated how the orientational order of $+1/2$ defects
persists in time and space. We find that within a single field of view
($2.3 \times 1.7$ mm) the axis of defect order does not change
appreciably over the entire sample lifetime (Fig. 2g). Therefore, we
used a motorized x-y stage to repeatedly scan centimeter-sized samples
every ten minutes, allowing us to measure long-range variations in
defect ordering. In such samples we measure nearly uniform
system-spanning orientational order (Fig. 2c, 2d). The largest active
nematic sample analyzed (5cm $\times$ 2cm) contained $\sim$20,000
defects, demonstrating that orientational order persists at scales
larger than 100 average defect spacings (Supplementary video 3). The
defect orientational order is a result of spontaneously broken
symmetry, and is not strongly influenced by the sample boundaries.  To
demonstrate this, we have confined active nematics in a circular
geometry, finding that defects form a single uniformly aligned domain,
rather than aligning with the boundaries (Supplementary Fig. 1).
Consistent with these observations, we additionally note that active
nematics in rectangular channels do not strongly favor either the long
or the short axis of the channel.

In simulations, $+1/2$ defects also attain system-spanning
orientational order (Fig. 3b, d). However, in contrast to the nematic
defect ordering observed in experiments, in the computational system
defects align with polar symmetry, which leads to their net transport
along the preferred direction. Possible reasons for this difference
are discussed below.

Tuning filament density controls the strength of the emergent defect
order and can even transform the system into an isotropic state. To
quantify the degree of defect ordering, we measure the 2D polar and
nematic order parameters, $P=\langle \cos (\psi - \bar{\psi} )
\rangle$ and $S = \langle \cos ( 2 [ \psi - \bar{\psi} ] ) \rangle$,
respectively, where $\psi$ is the orientation of a $+1/2$ defect and
$\bar{\psi}$ is the mean orientation of all defects in a given system
configuration. We find that thin nematic films (low MT concentration,
hence low retardance) have high defect nematic order, $S$; increasing
the film thickness (high MT concentration, high retardance) decreases
the magnitude of $S$ to the point where defects become effectively
isotropic (Fig. 2a-b, 4a). A similar effect is observed in simulations
when varying the particle density (area fraction); at the lowest
densities studied, defects have relatively strong alignment,
$P$. Increasing density induces a transition to an isotropic state
(Figs. 3a-b, 4b). Spatial correlation functions of these order
parameters demonstrate that in all experimental and computational
systems with measurable defect ordering, defect correlations are
system-spanning (Fig. 4c, 4d). Though the density of material is an
easily tuneable control parameter, it influences many material
properties, including the rate of energy dissipation, elastic
constants, and the efficiency with which active stresses are
transmitted through the material, all of which influence emergent
properties of the system. For example, defect density in experiments
decreases weakly with the MT film thickness, while in simulations it
increases with rod density (Supplementary Fig. 2). Additional studies
are required to relate the microscopic parameters to defect alignment.

In both experiment and simulation, we find that defect-ordered systems
also display nematic alignment of the constituent filaments (Fig. 2f,
3e).  The existence of such ordering in the presence of large numbers
of defects contrasts sharply with equilibrium systems, in which
defects reduce or eliminate nematic order.  Moreover, the ordering of
both defects and constituent rods decreases with system density (Fig
4a-b, Supplementary Fig. 6), unlike equilibrium lyotropic liquid
crystals, in which order increases with density. These contrasts
suggest that the orientational order in active nematics is driven by
non-equilibrium dynamics, and cannot be accounted for purely by
equilibrium-like alignment of the underlying material.

Although $+1/2$ defects are preferentially created perpendicular to
the local nematic field\cite{Shi2013}, they do not keep this
orientation but are continually reoriented by their interactions with
the local environment. A $+1/2$ defect moving through a distorted
nematic field will follow a curved path, always remaining
perpendicular to the field in front of it, leading to complex
meandering trajectories. Simultaneously, the passage of a $+1/2$
defect leaves behind a distortion in which the nematic field is
rotated $90^\circ$ from its previous orientation, creating a
topological structure which cannot relax except by the passage of
additional defects.  These distortions strongly affect defect motion
even after the defect which formed them has moved away or been
annihilated.  We therefore anticipate that a theory capable of
explaining the origin of defect ordering will need to simultaneously
account for both the defects and the underlying field.

Since our simulation model mimics the microscopic symmetries of the
experiments, it is notable that the two systems exhibit different
emergent symmetries. In experiments, +1/2 defects align with nematic
symmetry, and the direction of defect ordering is aligned with the
average MT direction.  In simulations, defects align with polar
symmetry, and the underlying rods are, on average, offset $90^\circ$
from the defects.  A number of distinctions between the two systems
may account for these differences. First, analysis of the
distributions of bend and splay distortions suggests that the
simulation model has a much higher bend modulus (see Supplementary
Methods). Second, the experimental system is subjected to hard-wall
boundary conditions, which preclude global polar ordering of motile
defects. Third, the simulation model is dry, whereas the experiments
experience hydrodynamic interactions. Finally, simulations investigate
a true 2D system, while in experiments MTs can pass over each other.
Further studies will be required to disentangle these effects.

In summary, our work demonstrates that transient, short-lived motile
defects can form higher-order dynamical phases with persistent
orientational order. The existence of such phases in both the
experimental and computational systems suggests that this is a generic
feature of active nematics.

\bibliographystyle{naturemag}
\bibliography{paper}

% Figures

\begin{figure}
  \centering\includegraphics[width=.75\linewidth]{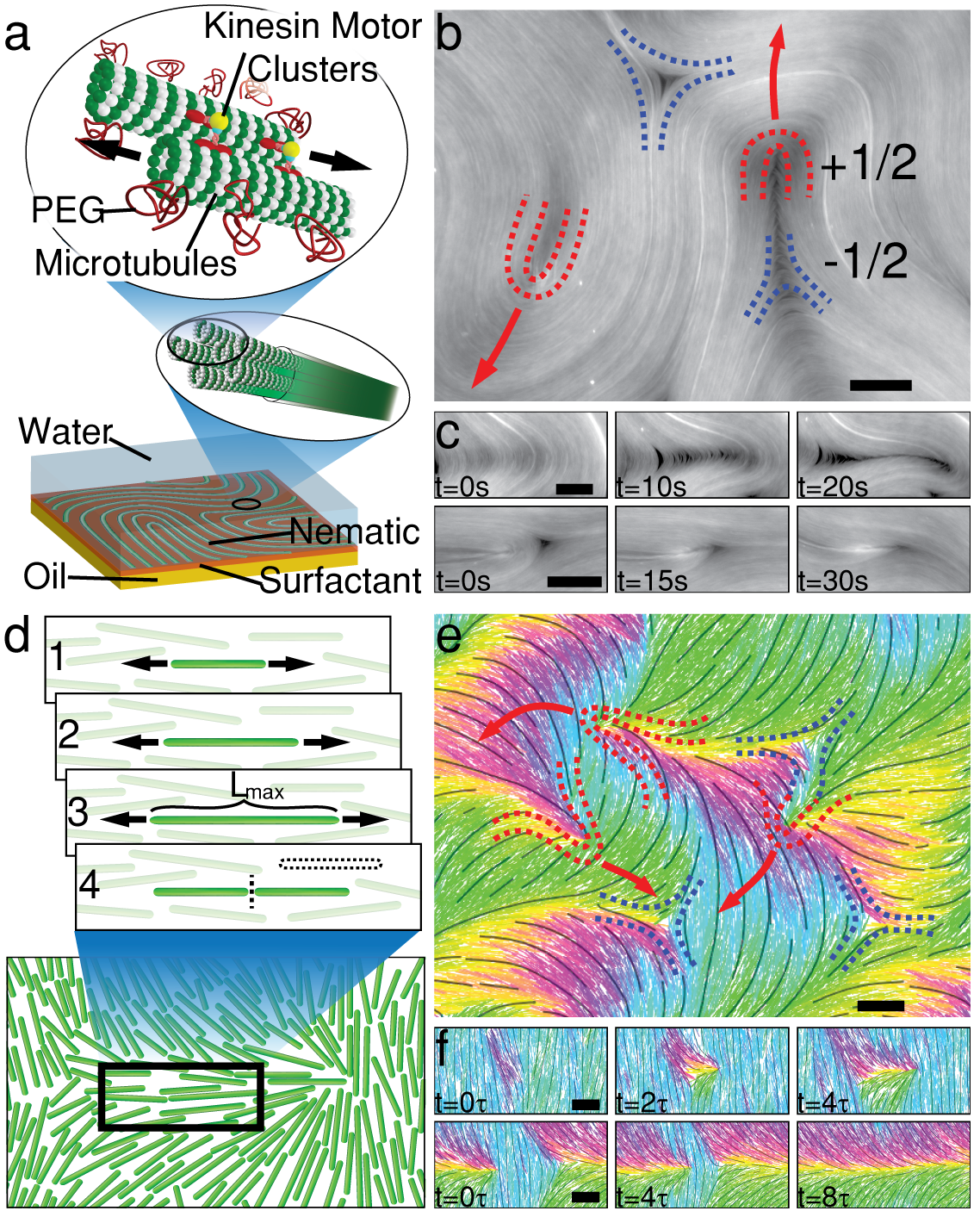}
  \caption{Overview of experimental and simulation systems.
    \textbf{a,} Microtubules (MTs) are bundled together by the
    depletion agent PEG. Kinesin clusters crosslink MTs and induce
    inter-filament sliding. Bundles are confined to a
    surfactant-stabilized oil-water interface, where they form a
    quasi-2D active nematic film. \textbf{b,} Fluorescence microscope
    image of a MT active nematic with defects of charge +1/2 (red) and
    -1/2 (blue). \textbf{c,} Image sequence illustrating the creation
    (top) and annihilation (bottom) of a defect pair. Scale bars
    50$\mu$m. \textbf{d,} Simulation microdynamics, consisting of hard
    rods which grow and split, reminiscent of the extension of MT
    bundles.  $L_{\text{max}}$ is the length at which a rod is
    split. \textbf{e,} Snapshot of simulated active nematic with
    marked defects. Rod colors indicate their orientations, and black
    streamlines guide the eye over the coarse-grained nematic
    field. \textbf{f,} Creation (top) and annihilation (bottom) events
    occur analogously to those in experiments. Scale bars $2 \Lmax$.}
\end{figure}

\begin{figure}
  \centering\includegraphics[width=.95\linewidth]{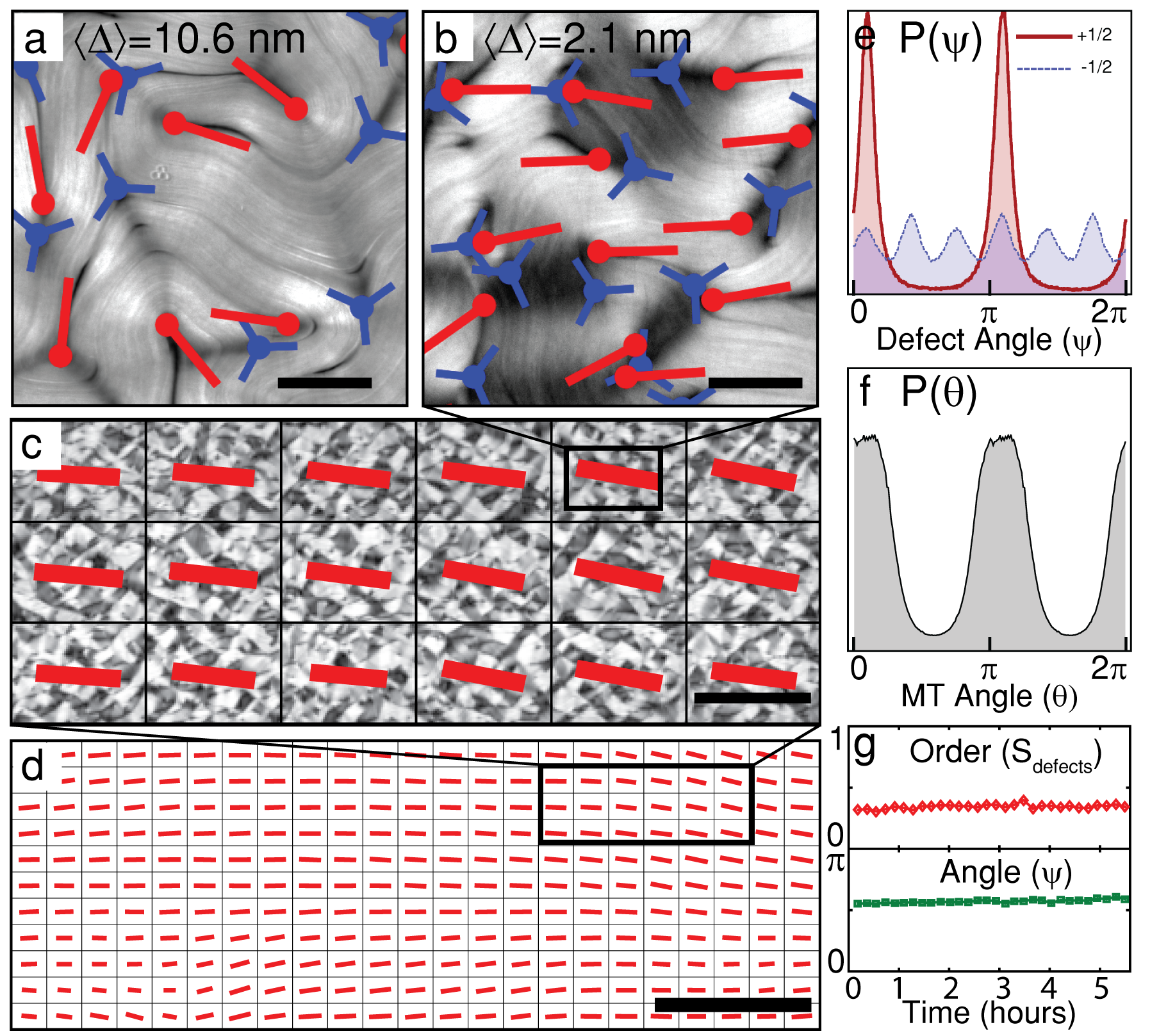}
  \caption{Defect-ordered phase in experiments. \textbf{a,} Retardance
    map of a thick MT film in the regime of weak defect alignment. Red
    and blue markers indicate locations and orientations of $+1/2$ and
    $-1/2$ defects. Scale bar 200$\mu$m. \textbf{b,} Thin MT film
    showing strong alignment of $+1/2$ defects. Scale bar
    200$\mu$m. \textbf{c,} Orientational order in a large active
    nematic sample. Each red bar's orientation and length indicates
    the mean direction and strength of defect alignment in one field
    of view. Scale bar 2mm. \textbf{d,} Defect alignment spans the
    largest samples studied (6cm $\times$ 2cm), containing
    $\sim$20,000 defects. Scale bar 10mm.  \textbf{e,} Normalized
    histogram of $+1/2$ (red) and $-1/2$ (blue) defect orientations,
    $P(\psi)$. \textbf{f,} MT orientation, $P(\theta)$, for the sample
    shown in panels b-d. Measurements of $P(\theta)$ from $0$ to $\pi$
    are duplicated from $\pi$ to $2 \pi$.  Both $P(\psi)$ and
    $P(\theta)$ show strong nematic ordering.  \textbf{g,} The
    preferred defect orientation (green) and magnitude of the order
    parameter (red) averaged over a field of view persists over the
    entire sample lifetime.}
\end{figure}

\begin{figure}
  \centering\includegraphics[width=.95\linewidth]{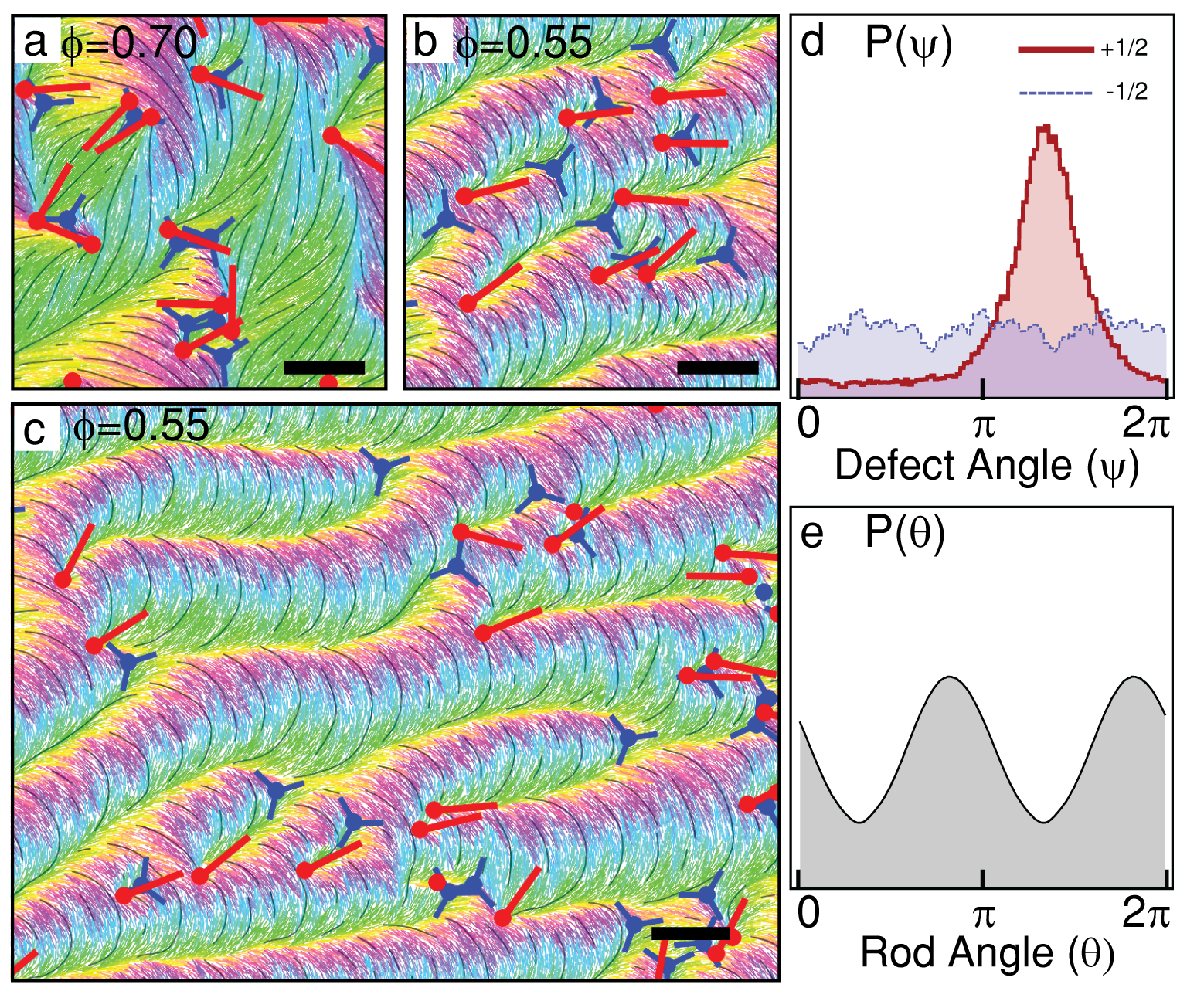}
  \caption{Defect-ordered phase in simulations. \textbf{a,} Snapshot
    of a high-area-fraction simulation in which $+1/2$ defects are not
    aligned. Red and blue markers indicate locations and orientations
    of $+1/2$ and $-1/2$ defects. Scale bar $5 \Lmax$. \textbf{b,} A
    low-area-fraction system in which defects show flocking-like polar
    alignment. Scale bar $5 \Lmax$. \textbf{c,} A large simulation
    with defects aligned over long distances. Scale bar $5
    \Lmax$. \textbf{d,} Normalized histogram of $+1/2$ (red) and
    $-1/2$ (blue) defect orientations, $P(\psi)$. \textbf{e,} Rod
    orientations $P(\theta)$ measured in the same sample. The former
    shows polar ordering, while the latter exhibits nematic ordering,
    and the preferred axes are offset by 90 degrees. Measurements of
    $P(\theta)$ from $0$ to $\pi$ are duplicated from $\pi$ to $2
    \pi$.}
\end{figure}

\begin{figure}
  \centering\includegraphics[width=.95\linewidth]{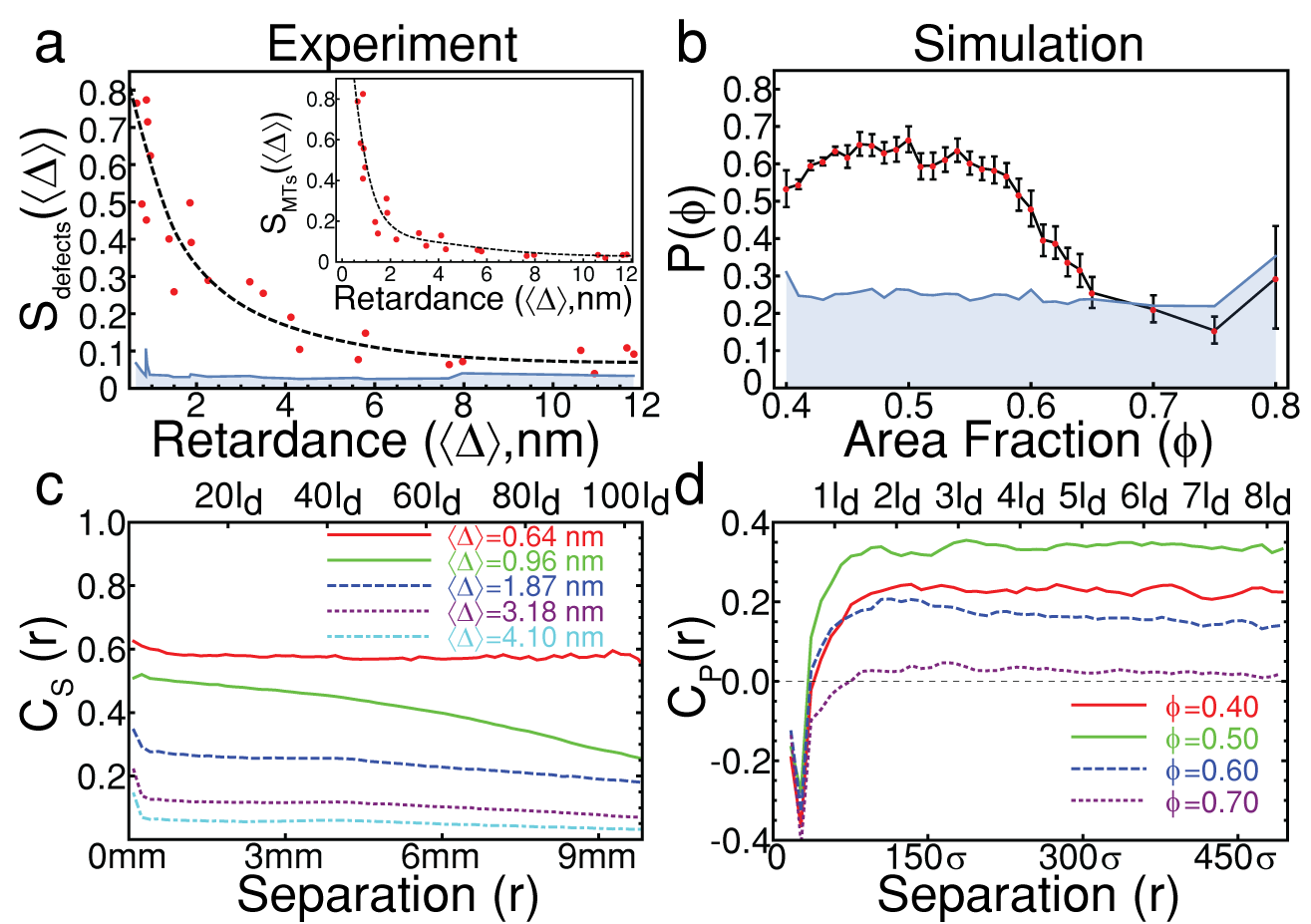}
  \caption{Quantitative measurements of defect alignment.  \textbf{a,}
    The defect nematic order parameter, $S$, decreases as a function
    of the MT film's retardance.  The blue shaded region represents
    the ``noise floor'' (see Supplementary Information).  Inset: The
    nematic order parameter of the underlying MT filaments, $S$, also
    decreases with the MT film's retardance.  \textbf{b,} The polar
    defect order parameter, $P$, showing a transition between ordered
    and isotropic regimes as a function of particle density. Error
    bars indicate a 90\% confidence interval computed by bootstrap
    methods (see Supplementary Information for details). \textbf{c,}
    The two-point nematic correlation of defect orientation $C_S(r) =
    \left\langle \cos 2 \left( \psi(\boldsymbol{r}) - \psi(0) \right)
    \right\rangle$ in MT films, which shows that orientational order
    persists over long distances. $l_d$ indicates the mean
    inter-defect spacing.  The magnitude of ordering falls as
    retardance increases. \textbf{d,} The polar correlation of defect
    orientation $C_P(r) = \left\langle \cos \left(
    \psi(\boldsymbol{r}) - \psi(0) \right) \right\rangle$ in
    simulations.  $\sigma$ indicates the width of a single rod (see
    Supplementary Information). At short ranges, defects tend to point
    in opposing directions, but beyond the first shell of neighbors,
    defects are likely to be aligned in the same direction.}
\end{figure}

\begin{addendum}
\item[Acknowledgments] The experimental portion of this study was
  primarily supported by Department of Energy, Office of Basic Energy Sciences
  through award DE-SC0010432TDD (SJD and ZD). The
  computational portion of this work (GSR, MFH, AB) was supported by
  NSF-MRSEC-1206146 and NSF-DMR-1149266. Computational resources were
  provided by the NSF through XSEDE (Stampede and Trestles) and the
  Brandeis HPCC, which is partially supported by the Brandeis MRSEC
  (NSF-MRSEC-1206146). We acknowledge use of a MRSEC optical facility
  that is supported by NSF-MRSEC-1206146.
  \item[Contributions] SJD and ZD conceived the experiments and GSR,
    AB and MFH conceived the simulations. SJD acquired experimental
    data. GSR performed computer simulations. SJD and GSR analyzed
    defect dynamics. SJD, GSR, AB, MFH and ZD wrote the paper. All
    authors revised the manuscript.
\item[Competing Interests] The authors declare that they have no
  competing financial interests.
\item[Correspondence] Correspondence and requests for materials should
  be addressed to M.F.H. \\ (hagan@brandeis.edu) or
  Z.D. (zdogic@brandeis.edu).
\end{addendum}

\end{document}